# Surface Waves Everywhere

M. I. Dyakonov

*Laboratoire Charles Coulomb, Université Montpellier, CNRS, France*

1. Introduction

By "surface waves" one means a special kind of waves that propagate at the interface between two different media. There exists a large variety of such waves, which are interesting on their own, and sometimes have also practical importance and technological applications. This article presents a brief and non-exhaustive review of this vast subject, designed as an introduction to the field for non-specialists. Similarities between surface waves of completely different origin will be outlined. I will be concerned with the physical picture and avoid math as strongly as possible (no differential equations and no boundary conditions), as well as many interesting details. Sometimes I will omit numerical coefficients. My own contributions to this field will be also presented.

It should be understood that the material in each section of this article is a subject of many books and hundreds, if not thousands, of journal publications, which the interested reader should address for more information.

2. Water waves

These are certainly the first kind of surface waves that mankind has encountered. However, our prehistoric ancestors did not yet realize that waves are characterized by the frequency $\omega$ and the wave vector $k = 2\pi/\lambda$, where $\lambda$ is the wavelength. The relation between $\omega$ and $k$ is called the *dispersion law*. We will now establish the dispersion laws for water waves from considerations of dimensionality only.

However, first we must distinguish between several kinds of water waves, differing in the nature of the return force that drives the oscillatory motion: this can be either *gravity* or *surface tension*.

Also we must distinguish between the cases of *deep* and *shallow* water. To do this, we must compare the water depth $h$ with some other length. There are two characteristic lengths in our problem: the wavelength $\lambda$ and the wave amplitude. If we are concerned with the linear regime, the amplitude is irrelevant (so long as it is small compared to the wavelength). Thus in the linear regime we must compare $h$ and $\lambda$, so that one has deep water when $h >> \lambda$ and shallow water if $h << \lambda$.

## 2.1 Gravitational waves

The return force is provided by gravity. In this case we have the following material for constructing the link between frequency $\omega$ with dimension $[s^{-1}]$ and the wave vector $k$ with dimension $[cm^{-1}]$: the freefall acceleration $g$ $[cm/s^2]$, and water depth $h$ [cm].

- *Deep-water waves* ($h \gg \lambda$).

In this case, the depth $h$ is irrelevant. Then the only possibility for satisfying the correct dimensionality is the following dispersion law: $\omega = (gk)^{1/2}$. The phase velocity $v = \omega/k$ is then given by $v \sim (g\lambda)^{1/2}$, so that longer wavelengths propagate faster than shorter ones. Beneath the surface, the water is perturbed on a depth $\sim \lambda$ (not the wave amplitude!), thus a diver does not care about the storm at the surface, so long as he stays at depths greater than the typical wavelength.

- *Shallow-water waves* ($h \ll \lambda$).

Now, independently of the wavelength the wave velocity is $v \sim (gh)^{1/2}$ and the dispersion law is linear: $\omega = vk$, as for sound waves. This formula gives a simple estimate for the time needed for a perturbation of the ocean surface caused by an earthquake to cross the Pacific and cause a tsunami on a distant coast.

The so-called "shallow earthquakes" have their focus up to 70 km deep in the Earth crust (say 50 km), which is much greater than the ocean depth (~3 km). Thus the ocean surface will be perturbed on a scale of 50 km and the resulting surface wave will be a *shallow water wave*, however strange it may sound when applied to the Pacific Ocean. This wave will propagate with a velocity $\sim (gh)^{1/2}$ where $h \sim 3$ km. This is about the velocity of an aircraft!

Note that gravity waves do not depend on the density of the liquid, because, as Galileo showed by dropping stones from the Pisa tower, the motion in the gravity field does not depend on the mass of the object.

## 2.2 Capillary waves

Here, the return force is due to surface tension $\sigma$. As before, we construct the dispersion law from dimensionality considerations only. This time our ingredients are: surface tension $\sigma$ $[g/s^2]$ and water density $\rho$ $[g/cm^3]$. Then only possible form of the dispersion law is: $\omega \sim (\sigma k^3/\rho)^{1/2}$, which leads to the wave velocity $v \sim (\sigma k/\rho)^{1/2} \sim (\sigma/\rho\lambda)^{1/2}$. Note, that in contrast to gravitational waves, now the wave velocity *decreases* with increasing wavelength.

Thus, generally the dependence of velocity on wavelength is nonmonotonic: there exists a minimal velocity $v_{MIN}$ at a wavelength $\lambda_{MIN}$ at which the roles of gravity and surface tension are roughly equal. For water, $\lambda_{MIN}$ is about 2 cm, and the corresponding velocity $v_{MIN}$ is about 30 cm/s.

## 2.3 Excitation of water waves by wind

The wind can excite water waves with phase velocities that are inferior to the wind velocity. The previous analysis shows that if the wind velocity is less than $v_{MIN}$, it cannot excite any waves at all. This is the rare case when the water surface is absolutely flat and mirror-like.

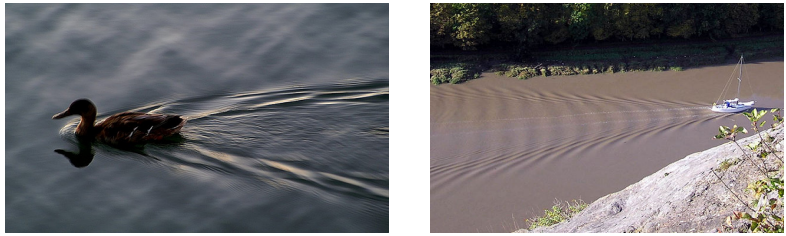

**Figure 1**. The V-shaped trail (the wake) behind a duck and a small boat.

On the other hand, when the wind speed exceeds $v_{MIN}$, it will excite both the long gravity waves and the short capillary waves; as a consequence we see long waves covered with capillary ripples.

### 2.4 Kelvin wake

A duck swimming in a pond leaves a V-shaped trail behind. The same wake exists for a small fishing boat and for an enormous aircraft carrier. The amazing fact is that the wake angle is universal: it is the same for the duck and the aircraft carrier, and it does not depend on the shape and speed of the vessel, nor on the density of the liquid (provided that the excited waves are deep water gravitational waves).

Lord Kelvin[1] showed that the wake angle θ is determined by the simple equation $\sin(\theta/2) = 1/3$, giving θ = 19.5°. This celebrated result is a direct consequence of the dispersion law for deep-water gravitational waves: $\omega = (gk)^{1/2}$.

### 2.5 Rogue waves

For many centuries, these waves, also known as freak waves, monster waves, killer waves, and extreme waves, have been considered as mythical. However, they are a real phenomenon, consisting in a sudden appearance, on the background of a mild ~5 meter wave pattern, of a huge ~30 meter single wave (the height of a 10-story building!), looking like a "wall of water" with a deep through in front.

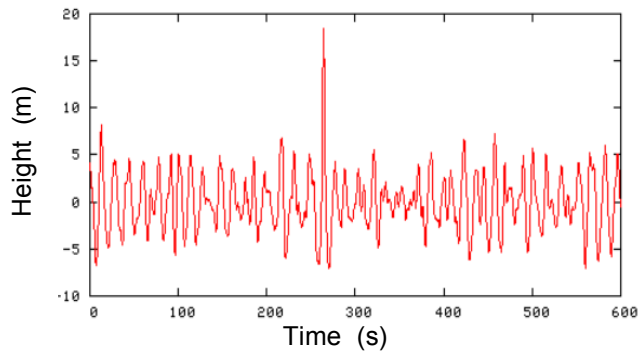

**Figure 2**. The "Draupner wave", a giant wave measured on New Year's Day 1995 in the North Sea, finally confirmed the existence of freak waves.[2]

The first scientific registration of rogue wave is presented in Fig. 2.[2] With increasing maritime traffic such extreme waves are seen more and more often.

So far, there does not exist any theoretical understanding of rogue waves, and they still remain a mystery. The problem can be formulated as that of the statistics of extremely rare events, which apparently is not Gaussian: large fluctuations occur more frequently than the conventional theory predicts. There are indications[3,4] that similar phenomena exist in optical and electrical noise.

3. Surface acoustic waves

In contrast to liquids and gases, in solids there are two types of sound: longitudinal and transverse, the speed of the latter being greater because of the contribution of shear stress. In 1885, Rayleigh predicted[5] that the longitudinal and transverse acoustic waves existing in the bulk of a solid can combine to produce a hybrid surface mode (the Rayleigh wave) with a velocity inferior to the velocities of both longitudinal and transverse sound (this is a necessary condition for their existence, otherwise they would excite bulk acoustic waves and disappear). This hybridisation makes it possible to satisfy the boundary conditions at the surface. The amplitude of the Rayleigh wave decreases exponentially away from the surface.

Such waves are used for nondestructive testing and in some electronic devices.

4. Surface plasma waves and polaritons

These are electromagnetic waves which exist when the dielectric constant is positive on one side of the interface and negative on the other side.[6] Negative dielectric constant $\varepsilon(\omega)$ is a well-known property of plasma below the so-called plasma frequency $\omega_P = 4\pi n e^2/m$, where $n$, $e$, and $m$ are the electron density, charge and mass, respectively. The dielectric constant of plasma is given by:

$$\varepsilon(\omega) = 1 - (\omega_P/\omega)^2 . \qquad (1)$$

Electromagnetic waves cannot propagate in a plasma if $\omega < \omega_P$. This explains the possibility of radio communication with the opposite side of the Earth, a fact that was puzzling at the dawn of the radio era. Later, it was understood that the ionosphere that acts as a mirror for frequencies below $\omega_P$. In contrast, for television one needs much higher frequencies (because of the necessity of having a broad band), which are greater than the characteristic plasma frequency in the ionosphere. This is why, until the invention of cable television, one had to build high towers with emitting antennas and television could be received only in areas within the horizon viewed from the top of the tower.

Surface plasma waves can propagate along the interface between media with positive and negative dielectric constant (*e.g.* air and metal). Similarly, surface electromagnetic waves (polaritons) may exist at the air/dielectric crystal interface, where ε < 0 for frequencies between the limiting frequencies of the optical and acoustical phonons.

## 5. Plasma waves in two-dimensional structures

There exist two distinct situations: an ungated two-dimensional electron gas (2DEG), in a quantum well or heterostructure, and a 2DEG with a metallic gate, like in a field-effect transistors (FET). The theoretical analysis of electromagnetic waves in such structures was done in Refs. 6–8, whereas the first experimental results were obtained in Refs. 9–11.

### 5.1 Ungated electron gas

The relevant properties of the 2D plasma are the electron concentration, $n$, the electron charge, $e$, and the electron effective mass, $m^*$. From dimensional considerations one obtains the following dispersion relation:

$$\omega(k) = (2\pi n e^2 k/m^* \varepsilon)^{1/2} . \qquad (2)$$

(Obviously, the $2\pi$ numerical factor cannot be guessed from dimensionality arguments. One needs some math to obtain this factor correctly).

Comparing this result with that of Section 2.1 one can notice a striking similarity with the dispersion law for *deep-water waves*! One of the consequences is that a phenomenon similar to the Kelvin wake should also exist in a 2DEG. Certainly, we cannot sail a boat in our heterostructure, however we could make some inclusion, or defect in the 2D gas and pass a current. Then, a static Kelvin wake with an angle of 19.5° will appear in the downstream direction and produce some measurable electrical disturbance at the lateral boundaries of the sample. So far such an experiment was never done.

### 5.2 Gated electron gas or FET

In addition to the electron concentration, in a gated 2DEG we now have another important parameter: the gate-to-channel capacitance per unit area $C_G$. The local electron charge in the channel $ne$ is related to the local gate-to-channel voltage $V_G$ by the plane capacitor formula: $ne = C_G(V_G - V_T)$, where $V_T$ is the threshold voltage. For gate voltages below $V_T$ there are no electrons in the channel. The difference $U = V_G - V_T$ is called *the gate voltage swing*.

Plasma waves in a FET consist in joint oscillations of local gate voltage and electron concentration in the channel. Dimensionality arguments lead to the existence of a characteristic velocity:

$$v = (ne^2/m^* C_G)^{1/2} = (eU/m^*) , \qquad (3)$$

and hence the dispersion law for plasma waves in a FET should be $\omega(k) = vk$.

Note again the interesting similarity with shallow water waves (or sound waves) described in Section 2.1. One has only to replace the *gravitational* energy per unit mass *gh* for shallow water waves by the *electrostatic energy* per unit mass $eU/m^*$ in the case of plasma waves in a FET!

This analogy has led Michael Shur and me to the prediction of plasma wave instability[12] in a FET for sufficiently high source-drain current provided the boundary conditions at the source and drain are asymmetric. A similar instability is responsible for the performance of wind musical instruments (the clarinet is asymmetric and no sound is produced unless one blows strongly enough).

## 6. Electronic surface states in solids

In a crystal, the periodic structure of the lattice results in the band structure of the electron energy spectrum and the states are described by Bloch wavefunctions. However the termination of the periodic structure at the boundary of the crystal will mix the wavefunctions belonging to different bands and this leads to the formation of specific electron surface states.

### 6.1 Tamm and Shockley states
In the early days of the quantum theory of solids, Tamm[13] and later Shockley[14] demonstrated the existence of such states within a model illustrated in Fig. 3. Tamm used the tight-binding approximation, while Shockley used the nearly free electron approximation. The physical nature of Tamm states is identical to that of Shockley states.

While being quite important as a matter of principle, these results do not have predictive value. In particular, the energy of Tamm-Shockley surface states crucially depends on the exact point chosen for termination of the periodic potential, which clearly is unphysical.

### 6.2 Surface states in HgTe quantum wells
It is well known that the electron spectrum in a quantum well consists of two-dimensional minibands. In a rectangular infinitely deep quantum well the textbook energy spectrum is given by $E_n(k) = \hbar^2\pi^2 n^2/2ma^2 + \hbar^2 k^2/2m$, where $n = 1, 2, 3…$, $a$ is the well width, $m$ is the electron mass, and $k$ is the in-plane electron momentum.

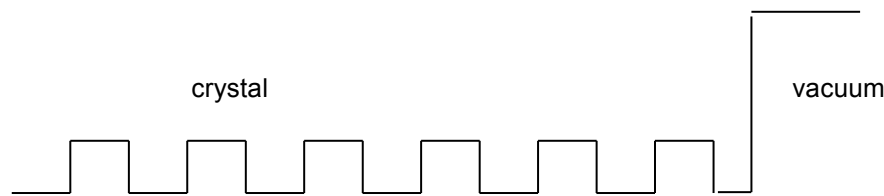

**Figure 3**. The electron potential energy in the vicinity of the boundary with vacuum.

Some 33 years ago, Khaetskii and I[15] became interested in understanding the energy spectrum in a quantum well formed by a gapless semiconductor, like HgTe. The bulk energy spectrum of this peculiar material is presented in Fig. 4(a). The conduction and the valence bands have the same origin as bands of light and heavy holes in Ge, GaAs, and other III-V semiconductors, in which there exist light and heavy holes with a degeneracy at $k = 0$. The specific feature of HgTe is that the light-hole band is inverted, becoming the *conduction* band. The hole is ~20 times heavier than the electron.

One could expect that in a quantum well the holes and the electrons would be quantized independently according to the above simple formula. This is indeed the case at $k = 0$, see Fig. 4(b). However, at finite $k$ we have a surprise: the first hole subband ($h1$) is inverted (it becomes electronic-like). The electron subbands ($e1$, $e2$, …) as well as other hole subbands ($h2$, $h3$, …) are "normal".

Moreover, while the wavefunction in the $h1$ band at $k = 0$ has a maximum in the middle of the well (which is normal), for large $k$ the maxima of the wavefunction appear near the well boundaries. This means the at large $k$ the $h1$ band is formed by *surface states*! If the Fermi level is located in this band, but below the $e1$ band, only the surface states should contribute to conductivity.

We also predicted that similar surface states should exist at the boundary of vacuum with a bulk HgTe crystal.[16] However, in this case the band of surface states is superimposed on the bulk band spectrum.

The subject was further explored in Refs. 17 and 18. In particular, Ref. 17 considered a quantum well of HgTe sandwiched between thick CdTe layers and made the first prediction of "time-reversal symmetry protected edge states", which were confirmed experimentally exactly 20 years later.[19]

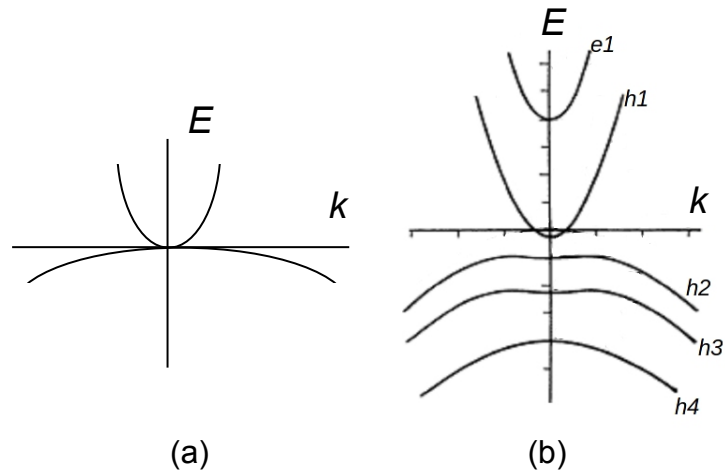

(a)          (b)

**Figure 4**. The bulk spectrum of the gapless semiconductor HgTe (a) and the two-dimensional spectrum in a HgTe quantum well[15] (b).

## 6.2 Topological insulators

The theoretical work of Pankratov, Pakhomov, and Volkov[17] was the first prediction belonging to the now very popular field of *topological insulators*. The exciting property is that in such materials the *surface* states (in 3D materials) or the *edge* states in 2D sandwich-like quantum wells are *time-reversal symmetry protected*, which means that they are immune to potential scattering and spin-orbit interaction (but can be destroyed by scattering by magnetic impurities, *i.e.* by spin scattering).

The 3D topological insulator was predicted by Fu and Kane[19] for binary compounds involving bismuth. The predicted symmetry-protected surface states were discovered in BiSb.[20] The surface states of a 3D topological insulator is a new type of a 2DEG where the electron's spin is locked to its linear momentum.

There is a huge recent literature on this subject so we will not dwell on it further.

## 7. Dyakonov surface waves (DSWs)

Inspired by Rayleigh results for surface acoustic waves,[5] some time ago[21] I started a search for surface waves in another situation, where there exist two kinds of waves in the bulk: the ordinary and extraordinary electromagnetic waves in a birefringent crystal with eigenvalues of the dielectric tensor $\varepsilon_{\|}$ and $\varepsilon_{\perp}$. (It is interesting to note that all the important facts about optics of anisotropic crystals were established by Fresnel[22] long before Maxwell's equations were written). The simplest configuration in which these new surface waves can exist is presented in Fig. 5.

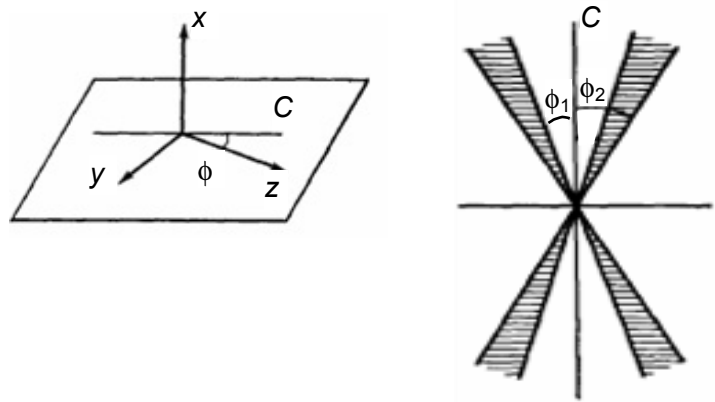

**Figure 5**. (a) The interface between an isotropic medium ($x > 0$) and a birefringent crystal ($x < 0$). The optical axis of the crystal is denoted by *C*. The direction of the surface wave phase velocity (*z*) makes an angle *φ* with the optical axis. (b) The angular intervals where the surface waves can propagate.

In order to describe waves decaying away from the interface ($x = 0$), the $x$-components of their wavevectors should be imaginary: $ik_1$, $-ik_2$, and $-ik_3$ (for the wave in the isotropic media at $x > 0$, the extraordinary wave in the crystal at $x < 0$, and for the ordinary wave in the crystal, respectively). The dispersion laws for the three waves are:

$$q^2 - k_1^2 = \varepsilon ,$$
$$(q^2\sin^2\phi - k_2^2)/\varepsilon_\| + (q^2\cos^2\phi)/\varepsilon_\perp = 1 , \quad (4)$$
$$q^2 - k_1^2 = \varepsilon_\perp ,$$

where all the wavevectors are dimensionless (written in units of $\omega/c$).

These equations should be complemented by the boundary conditions requiring the continuity of the tangential components of the electric and magnetic fields at the interface. After some tedious algebra, one can obtain the fourth equation:[23]

$$(k_1 + k_2)(k_1 + k_3)(\varepsilon k_3 + \varepsilon_\perp k_2) = (\varepsilon_\| - \varepsilon)(\varepsilon - \varepsilon_\perp)k_3 . \quad (5)$$

Now, we can determine the four unknowns: $k_1$, $k_2$, $k_3$, and $q$, which can be done only numerically. However, from the above equation one can see that a solution can exist only if $\varepsilon_\| > \varepsilon > \varepsilon_\perp$ or $\varepsilon_\perp > \varepsilon > \varepsilon_\|$. Further analysis shows that only the first case is acceptable, thus the condition for existence of DSW is:

$$\varepsilon_\| > \varepsilon > \varepsilon_\perp , \quad (6)$$

which means that the birefringent crystal must be *positive* ($\varepsilon_\| > \varepsilon_\perp$).

Another possibility was considered by Averkiev and me.[24] Consider two identical positive birefringent crystals with optical axes parallel to their surfaces. Put one upon the other and make some angle $\alpha$ between their optical axes $C_1$ and $C_2$, as shown in Fig. 6(a). For such a configuration, we found that surface waves can exist within two sectors around the bisectrixes of the two angles formed by $C_1$ and $C_2$, as shown in Fig. 6(b).

The previously known electromagnetic surface waves (surface plasmons discussed in Section 4) exist under the condition that the permittivity of one of the materials forming the interface is negative, while the other one is positive. In contrast, the DSWs can propagate when both materials are transparent; hence they are virtually lossless, which is their most attractive property. A large number of theoretical works appeared discussing these, or similar, waves under various conditions, in particular, in metamaterials,[25] see the review in Ref. 26.

The first experimental observation of DSW was reported only in 2009 by Takayama and co-workers,[27,28] using the Otto-Kretchmann configuration and an index-matching fluid as an isotropic medium satisfying the condition $\varepsilon_\| > \varepsilon > \varepsilon_\perp$. The cutoff propagation angle was measured as a function of the permittivity of the fluid, see Fig. 7.

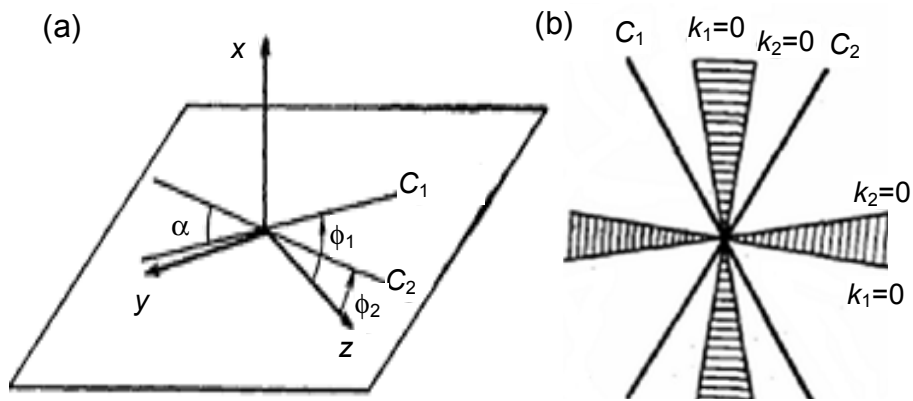

**Figure 6**. Surface electromagnetic waves at the interface between two identical birefringent crystals with optical axes $C_1$ and $C_2$ parallel to the interface and making an angle $\alpha$ between them: (a) geometry and (b) the angular intervals where surface waves can propagate.[21]

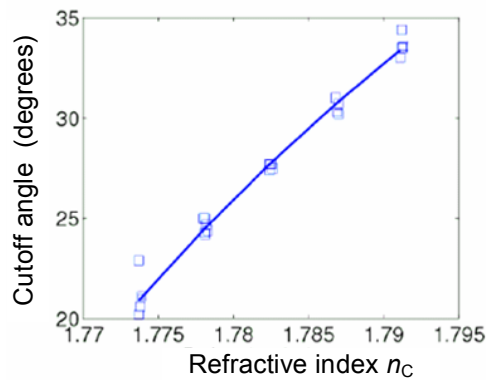

**Figure 7**. The measured upper cutoff angle as a function of the refractive index of the matching fluid.[27] Line – theoretical, points – experimental. Note the extremely fine tuning.

It is believed that the extreme sensitivity of DSWs to anisotropy, and thereby to stress, along with their low-loss (long-range) character render them particularly attractive for enabling high sensitivity tactile and ultrasonic sensing for next-generation high-speed transduction and read-out technologies.

This subject continues to be of considerable interest, recent work can be found in Refs. 29–34.

**References**

1. Lord Kelvin (W. Thomson), "Deep sea ship-waves," *Math. Phys. Papers* **4**, 303 (1910).
2. See *en.wikipedia.org/wiki/Draupner_wave*
3. D. R. Ropers, C. Koonath, and P. Jalali, "Optical rogue waves," *Nature* **450**, 1054 (2007).
4. X. Hu, P. Du, and C.-K. Cheng, "Exploring the rogue wave phenomenon in 3D power distribution networks," *Proc. IEEE 19th Conf. Electrical Perf. Electronic Packaging Syst.* (2010), pp. 57–60.
5. Lord Rayleigh, "On waves propagated along the plane surface of an elastic solid," (1885), see *http://plms.oxfordjournals.org/content/s1-17/1/4.full.pdf*
6. R. H. Ritchie, "Plasma losses by fast electrons in thin films," *Phys. Rev.* **106**, 874 (1957).
7. F. Stern, "Polarizability of a two-dimensional electron gas," *Phys. Rev. Lett.* **18**, 546 (1967)
8. M. Nakayama, "Theory of surface waves coupled to surface carriers," *J. Phys. Soc. Japan* **36**, 393 (1974).
9. A. Eguiluz, T. K. Lee, J. J. Quinn, and K. W. Chiu. "Interface excitations in metal-insulator-semiconductor structures," *Phys. Rev. B* **11**, 4989 (1975).
10. S. J. Allen, D. C. Tsui, and R. A. Logan, "Observation of the two-dimensional plasmon in silicon inversion layers," *Phys. Rev. Lett.* **38**, 980 (1977).
11. D. C. Tsui, E. Gornik, and R. A. Logan, "Far infrared emission from plasma oscillations of Si inversion layers," *Solid State Commun.* **35**, 875 (1980).
12. M. I. Dyakonov and M. S. Shur, "Shallow water analogy for a ballistic field effect transistor: new mechanism of plasma wave generation by dc current," *Phys. Rev. Lett.* **71**, 2465 (1993).
13. N. I. Tamm, "On the possible bound states of electrons on a crystal surface," *Phys. Z. Soviet Union* **1**, 733 (1932).
14. W. Shockley, "On the surface states associated with a periodic potential," *Phys. Rev.* **56**, 317 (1939).
15. M. I. Dyakonov and A. V. Khaetskii, "Size quantization of the holes in a semiconductor with a complicated valence band and of the carriers in a gapless semiconductor," *Sov. Phys. JETP* **55**, 917 (1982).
16. M. I. Dyakonov and A. V. Khaetskii, "Surface states in a gapless semiconductor," *JETP Lett.* **33** 110**,** (1981)
17. O. A. Pankratov, S. V. Pakhomov, and B. A. Volkov, "Supersymmetry in heterojunctions: Band-inverting contact on the basis of $Pb_{1-x}Sn_xTe$ and $Hg_{1-x}Cd_xTe$," *Solid State Commun.* **61**, 93 (1987).
18. I. G. Gerchikov and A. V. Subashiev, "Interface states in subband structure of semiconductor quantum wells," *phys. stat. sol. (b)* **160**, 443 (1990).
19. M. König, S. Wiedmann, C. Brüne, *et al.*, "Quantum spin Hall insulator state in HgTe quantum wells," *Science* **318**, 766 (2007).
20. L. Fu and C. L. Kane, "Topological insulators with inversion symmetry," *Phys. Rev. B* **76**, 045302 (2007).

21. D. Hsieh, A. L Dong Qian, Y. X. Wray, R. C. Yusan Hor, and H. M. Zahid, "A topological Dirac insulator in a quantum spin Hall phase," *Nature* **452**, 970 (2008).
22. M. I. Dyakonov, "New type of electromagnetic waves propagating at an interface," *Sov. Phys. JETP* **67**, 714 (1988).
23. A.-J. Fresnel, *Œuvres Complètes*, vol. 1, Paris: Imprimerie impériale, 1868.
24. N. S. Averkiev and M. I. Dyakonov, "Electromagnetic waves localized at the interface of transparent anisotropic media," *Optics Spectroscopy (USSR)* **68**, 653 (1990).
25. D. Artigas and L. Torner, "Dyakonov surface waves in photonic metamaterials," *Phys. Rev. Lett.* **94**, 013901 (2005)
26. O. Takayama, L. C. Crassovan, D. Mihalache, and L. Torner, "Dyakonov surface waves: A review," *Electromagnetics* **28**, 126 (2008).
27. O. Takayama, L. Crassovan, D. Artigas, and L. Torner, "Observation of Dyakonov surface waves," *Phys. Rev. Lett.* **102**, 043903 (2009).
28. L. Torner, D. Artigas, and O. Takayama, "Dyakonov surface waves," *Optics Photonics News* **20**, 25 (2009).
29. Z. Jacob and E. Narimanov, "Optical hyperspace for plasmons: Dyakonov states in metamaterials," *Appl. Phys. Lett*. **93**, 221109 (2008).
30. O. Takayama, D. Artigas, and L. Torner, "Practical dyakonons," *Optics Lett.* **37**, 20 (2012) .
31. O. Takayama, D. Artigas, and L. Torner, "Coupling plasmons and dyakonons," *Optics Lett.* **37**, 1983 (2012)
32. C. J. Zapata-Rodriguez, J. J. Miret, S. Vukovič, and Z. Jakšič, "Dyakonons in hyperbolic metamaterials,", *Photonics Lett. Poland* **5**, 63 (2013).
33. M. A. Noginov, "Steering Dyakonov-like waves," *Nature Nanotechnol.* **9**, 414 (2014).
34. O. Takayama, D. Artigas, and L. Torner, "Lossless directional guiding of light in dielectric nanosheets using Dyakonov surface waves," *Nature Nanotechnol.* **9**, 419 (2014).